\documentclass{aastex631}

\usepackage{graphicx}
\usepackage{amsmath}
\usepackage{amssymb}
\usepackage{color}
\usepackage{hyperref}
\usepackage{multirow}
\usepackage{booktabs}
\usepackage{array}

\usepackage{soul}

\newcommand{\pip}[1]{{\textcolor{black}{\texttt{Muphoten}}}}
\newcommand{\ddpip}[1]{{\textcolor{black}{\url{https://gitlab.in2p3.fr/icare/muphoten/}}}}

\begin{document}

\title{Muphoten: A Multiband Photometry Tool for Telescope Networks}

\author{P. A. Duverne}
\affiliation{Université Paris-Saclay, CNRS/IN2P3, Orsay, France}

\author{S. Antier}
\affiliation{Universit\'e Côte d'azur, CNRS, Artemis, F 06304 NICE Cedex 4, France}

\author{S. Basa}
\affiliation{Laboratoire d'Astrophysique de Marseille, UMR 7326, CNRS, Universit\'e d'Aix Marseille, CNRS, CNES, LAM, IPhU, 38, rue Fr\'ed\'eric Joliot-Curie,Marseille, France}

\author{D. Corre}
\affiliation{Université Paris-Saclay, CNRS/IN2P3, Orsay, France}
\affiliation{Sorbonne Université, CNRS, UMR 7095, Institut d’Astrophysique de Paris, 98 bis boulevard Arago, 75014 Paris, France}

\author{M. W. Coughlin}
\affiliation{School of Physics and Astronomy, University of Minnesota, Minneapolis, MN 55455, USA}

\author{A. V. Filippenko}
\affiliation{Department of Astronomy, University of California, Berkeley, CA 94720-3411, USA}
\affiliation{Miller Institute for Basic Research in Science, University of California, Berkeley, CA 94720, USA}

\author{A. Klotz}
\affiliation{IRAP, Universit\'e de Toulouse, CNRS, UPS, 14 Avenue Edouard Belin, F-31400 Toulouse, France}
\affiliation{Universit\'e Paul Sabatier Toulouse III, Universit\'e de Toulouse, 118 route de Narbonne, 31400 Toulouse, France}

\author{P. Hello}
\affiliation{Université Paris-Saclay, CNRS/IN2P3, Orsay, France}

\author{W. Zheng}
\affiliation{Department of Astronomy, University of California, Berkeley, CA 94720-3411, USA}



\begin{abstract}

The early and complete temporal characterization of optical, fast, transient sources requires continuous and multiband observations over different timescales (hours to months). For time-domain astronomy, using several telescopes to analyze single objects is the usual method, allowing the acquisition of highly sampled light curves. Taking a series of images each night helps to construct an uninterrupted chain of observations with a high cadence and low duty cycle. Speed is paramount, especially at early times, in order to capture early features in the light curve that help determine the nature of the observed transients and assess their astrophysical properties. However, the problem of rapidly extracting source properties (temporal and color evolution) with a heterogeneous dataset remains.
Consequently, we present \pip~, a general and fast-computation photometric pipeline able to address these constraints. It is suitable for extracting transient brightness over multitelescope and multiband networks to create a single homogeneous photometric time series. We show the performance of \pip~ with observations of the optical transient SN\,2018cow (from June 2018 to July 2018), monitored by the GRANDMA network and with the publicly available data of the Liverpool Telescope.

\end{abstract}


\section{Introduction}
\label{intro}
Recent years have seen the rise of multimessenger astronomy (MMA), which aims to provide follow-up observations of astrophysical events through joint observations of different messengers: photons, gravitational waves (GW), and high-energy particles.
Among these events, optical counterparts to gravitational-wave sources are of particular interest --- especially kilonovae, as they bring insights to various domains, including astrophysics, nuclear physics, and cosmology. So far, one kilonova has been confidently identified after the merger of binary neutron stars detected by the LIGO and Virgo Collaborations, GW\,170817 \citep{gw17}. This event launched a follow-up campaign \citep{MM17,kilo17,follow17} with ground-based telescopes from 70 observatories that confirmed the expected \citep{10.1093/mnras/stv721, Metzger2017} main observational constraints of kilonovae: they are faint ($\sim -$16\,mag at peak brightness) and rapidly evolving $\sim 0.5$\,mag\,day$^{-1}$ transients, providing a narrow time window for observations \citep{follow17}. 

However, follow-up observations triggered by gravitational waves bring numerous constraints. The main one is to deal with large sky areas ($\sim 100$--1000\,deg$^2$) that complicate the discovery and subsequent monitoring. Consequently, follow-up observations require much observing time. In addition, to have useful constraints on kilonova models, very low latencies must be achieved between the trigger detection and the first observations \citep{Arcavi_2018}. In order to meet these requirements, several networks of telescopes were set up. During the third observing run of LIGO and Virgo (April 2019 -- March 2020), the Global Rapid Advanced Network Devoted to the Multi-messenger Addicts (GRANDMA; \citealt{mamie1,mamie2}), Global Relay of Observatories Watching Transients Happen (GROWTH; \citealt{growth}), Gravitational wave Optical Transient Observer (GOTO; \citealt{GOTO}), and Mobile Astronomical System of TElescope Robots (MASTER; \citealt{master}) performed regular follow-up observations of gravitational-wave alerts.

Although using a network to detect, identify, and monitor a transient helps address the above constraints of, it also creates specific challenges. Several instruments observing simultaneously will produce hundreds to thousands of images. Yet, they need to be quickly reduced in order to know whether there is a source candidate in the image or if a candidate is consistent with a kilonova. Moreover, these images will have a range of quality because of the instruments' heterogeneity in the network. Furthermore, the rapid evolution of the transient precludes waiting for optimal observing conditions. Hence, the images need to be treated with a standardized procedure providing a consistent intercalibration for all the instruments to have rapid and reliable photometry estimates.

The GRANDMA network is composed of $\sim 30$ instruments, and it has a citizen-science program called KilonovaCatcher (see \citealt{mamie2}), for which tens of additional amateur astronomers may provide images. Hence, taking into account the various characteristics of the instruments, the data quality, and the observing conditions requires a dedicated method to provide consistent photometry.
To have a network able to perform low-latency observations, it is helpful to provide early photometric information about a newly detected transient. Eventually, these data can be used to determine whether a transient is worth being monitored based on some basic criterion, such as the decay rate or color evolution, before any spectrum of the transient is acquired.

In this paper, we present a publicly available photometry tool, Muphoten\footnote{\href{https://gitlab.in2p3.fr/icare/MUPHOTEN}{https://gitlab.in2p3.fr/icare/MUPHOTEN}}, that can be used by networks that monitor astrophysical objects with heterogeneous instruments. In Section~\ref{pipeline}, we describe the details of the image analysis. Section~\ref{application} presents the results of follow-up observations of AT\,2018cow by five instruments: the 50\,cm Initiation à la Recherche en astronomIe pour les Scolaires (IRiS)\footnote{\href{http://IRiS.lam.fr/}{http://IRiS.lam.fr/}}, the 0.76\,m Katzman Automatic Imaging Telescope (KAIT; \citealt{kaitpap}), and the 0.25\,m TAROT Chile \citep{tarot}. The last three instruments are members of the GRANDMA consortium and the images of SN\,2018cow are previously unpublished. The images we used also include the public data from the 2\,m Liverpool Telescope (LT; \citealt{lt}) and the Kitt Peak 2.1\,m telescope with the Kitt Peak EMCCD Demonstrator (KPED; \citealt{kped}).

\section{\textbf{Methodology}}
\label{pipeline}
\pip~ is a Python-based package that aims to standardize the analysis of images acquired by heterogeneous instruments. This method is designed to reduce images produced by several telescopes operating in various observing conditions and with different filter sets. It returns the apparent magnitude of sources whose WCS-coordinate positions have been specified by the user through the input file. 
The code is based on the public Python libraries \texttt{Photutils}\footnote{\label{photu} https://photutils.readthedocs.io/} \cite{phot} and \texttt{Astroquery}\footnote{\label{astr}https://astroquery.readthedocs.io/en/latest/} \cite{astroquery}, and it uses the publicly available software packages {\tt Swarp} \cite{swarp}, \texttt{PSFex} \cite{psfex}  and \texttt{HOTPANTS}\footnote{\label{hot}https://github.com/acbecker/hotpants} \cite{hotpants}. 

\begin{figure}[htbp]
  \centering
  \includegraphics[width=0.60\textwidth]{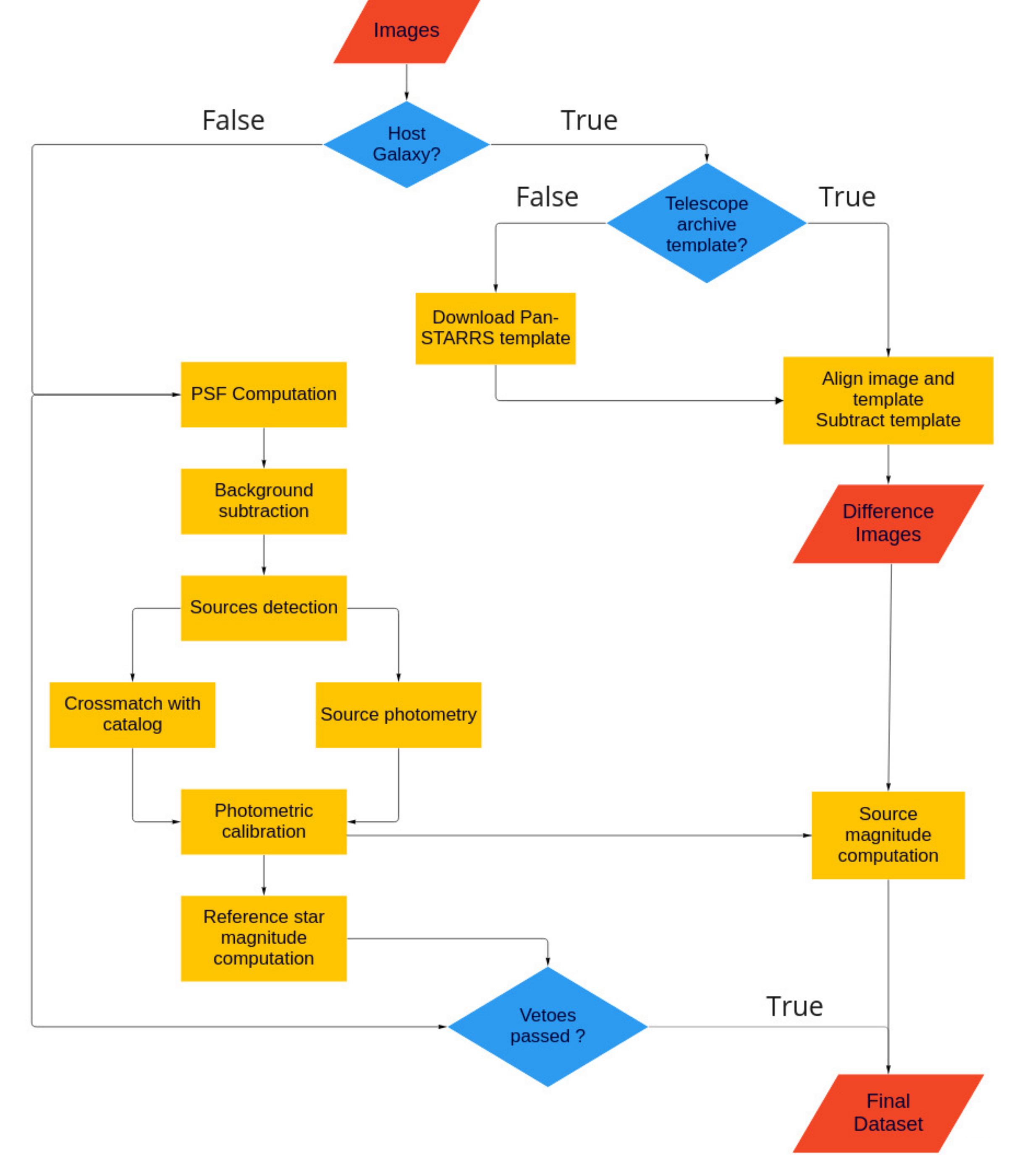}
  \caption{Muphoten flowchart.}
  \label{scheme}
\end{figure}

As shown in Figure~\ref{scheme}, the main steps of \pip~ are as follows.
\begin{itemize}
    \item The pipeline begins with a collection of pre-processed astronomical images of astrophysical sources (i.e., already dark, bias, and flat-field corrected). We require WCS-based astrometry to be available in the FITS headers, estimated at a $3''$ precision for the crossmatch with public surveys in the next steps.
    \item Subtraction of a \textit{template image} without the source to remove potential host-galaxy flux. The template can either be downloaded by Muphoten from the Pan-STARRS \cite{pssurvey} archive or provided by the user. Image subtraction is not necessary for a hostless source.
    \item Selection of the reference catalog among the four available: Pan-STARRS \cite{pssurvey}, the Sloan Sky Digital Survey \cite{sdsssurvey}, {\it Gaia} \cite{gaiasurvey}, and USNO-B1\cite{usnosurvey}.
    \item Estimation and subtraction of the background level. 
    \item Detection of all sources and computation of their instrumental magnitudes:
    \begin{equation}
        M_{\rm ins} = -2.5\, {\log}_{10}(N_{\rm \mathbf{ADU}}). 
    \end{equation}
    Currently, there are three available options: isophotal aperture, fixed aperture, and Kron photometry \citep{kron}.
    \item Crossmatching of the detected sources with the reference catalog.
    \item Photometric calibration by fitting the relation between the instrumental magnitude of the detected sources and their magnitude in the reference catalog.
    \item Computation of the magnitude for the source of interest using the previously obtained calibration.
    \item Identification and rejection of poor-quality images using two vetoes. The first is based on the point-spread function (PSF), and the second is based on the light curve of a known star in the field of view.
\end{itemize}
Each step is detailed below.

\begin{figure}[htbp]
 \includegraphics[width=1.0\textwidth]{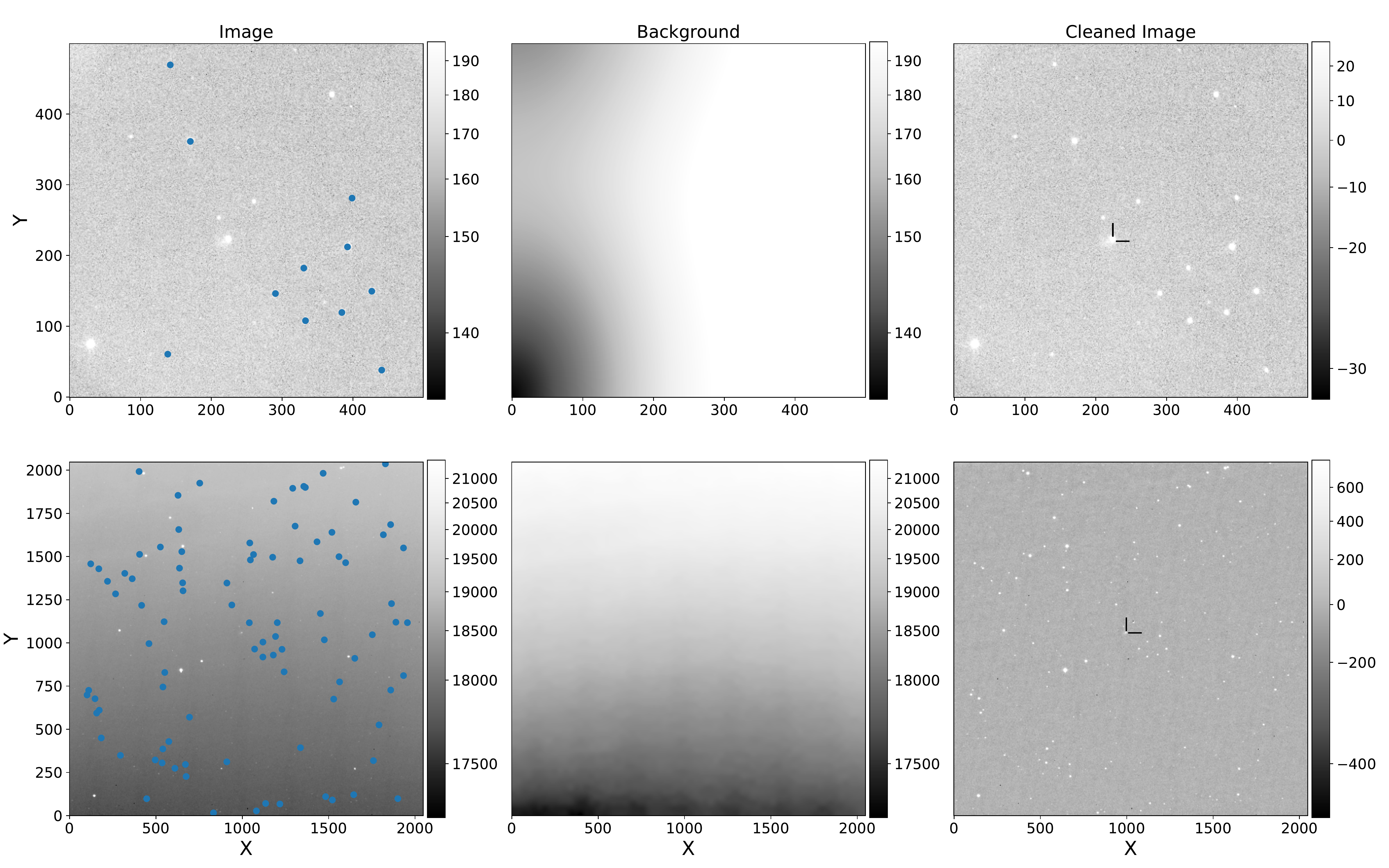}
  \caption{Example of images corresponding to the SN\,2018cow observational campaign taken by telescopes KAIT (upper panel) and IRiS (bottom panel). 
  Blue dots on the right are the sources detected and crossmatched with the reference catalog to obtain the photometric calibration. The center column shows the background images subtracted from science images to obtain the cleaned images on the right column. SN\,2018cow is indicated with black markers.}
  \label{images}
\end{figure}

\subsection{Host-Galaxy Subtraction}
\label{sub}
The host-galaxy flux must be removed if the source of interest is located in a galaxy. This is done by subtracting the template image in which the source is not visible, with {\tt HOTPANTS} \citep{hotpants}. The template image can be taken from the telescope archive or built with Pan-STARRS images.For the latter case, the template is constructed to match the observed field of view with a mosaic of Pan-STARRS stacked images \citep{ps1images}. When the image filter is not a Pan-STARRS one (e.g., Johnson-Cousins or Clear image), the closest filter of the analyzed image is used. It corresponds to $g_{PS1}$ for the $B$, $V$, and clear images, $r_{PS1}$ for $R_c$, and $i_{PS1}$ for $I_c$. Otherwise, the template can be provided by the user. For both cases, the analyzed image and the template are aligned with {\tt Swarp} \citep{swarp}. If necessary, the template image is also resampled to the analyzed image sampling with {\tt Swarp}. The difference image is then background subtracted, and the instrumental magnitude of the transient is computed in the same way as for the photometric calibration.

\subsection{Selection of a Reference Catalog}
\label{cata}
We employ reference catalogs to provide reference magnitudes for the sources in the image. \pip{}~uses four different optical catalogs: Pan-STARRS \cite{pssurvey}, SDSS \cite{sdsssurvey}), {\it Gaia} \cite{gaiasurvey}, and USNO-B1 \cite{usnosurvey}.

Pan-STARRS is the default choice as it provides photometry in five bands ($g_{\rm PS}$, $r_{\rm PS}$, $i_{\rm PS}$, $z_{\rm PS}$, and $y_{\rm PS}$) down to $\sim 21.5$--22\,mag and covers  all of the sky north of declination $-30^{\circ}$. We choose SDSS for images acquired in an ultraviolet filter ($U$ or $u'$ band), as Pan-STARRS does not cover this range. If the two previous options are not available, we select {\it Gaia}, which provides all-sky photometry in three bands ($G_{\rm BP}$, $G_{\rm RP}$, and $G$) covering optical wavelengths. This survey cannot be used for the $U$, $B$, $u'$, and $z'$ bands. If the three previous options cannot be used, we employ the USNO-B1 catalog with two bands: $B1$ and $R1$.
For {\it Gaia} and USNO that do not have standard filters, the magnitudes are transformed via the equations presented in Table~\ref{tab:filter} to match the analyzed image filter.

\begin{deluxetable}{c|c}[htbp]
    \label{tab:filter}
    \tablewidth{0.70\textwidth}
    \tablecaption{Temporal properties of the Cow derived with the data produced by \pip~. All of the values are consistent with the published literature about this transient.}
    \tablehead{ \colhead{\textbf{Relation to go from {\it Gaia} to SDSS DR 12 (\citealt{gaiatransfo})}} & \colhead{\textbf{Error}}  \\
                \colhead{-} & \colhead{[mag]}}
    \startdata
        $G - r = -0.049465\,(G_{\rm BP}-G_{\rm RP})^3 - 0.027464\,(G_{\rm BP}-G_{\rm RP})^2 + 0.24662\,(G_{\rm BP}- G_{\rm RP}) - 0.12879$  & 0.066739 \\
        $G - i = -0.10141\,(G_{\rm BP}-G_{\rm RP})^2 + 0.64728\,(G_{\rm BP}-G_{\rm RP}) - 0.29676$ & 0.098957 \\
        $G - g = 0.021349\,(G_{\rm BP}-G_{\rm RP})^3 - 0.25171\,(G_{\rm BP}-G_{\rm RP})^2 - 0.46245\,(G_{\rm BP}- G_{\rm RP}) + 0.13518 $ & 0.16497 \\
    \hline
    \hline
        \textbf{Relation to go from SDSS to Johnson-Cousins (\citealt{sdsstransfo_paper})} (for Pop I stars) & \textbf{Error} \\
    \hline
        $B - g = 0.312\,(g-r) + 0.219$ & 0.0036 \\
        $V - g = -0.573\,(g-r) - 0.016$ & 0.0028 \\
        $R_c - r = -0.257\,(r-i) + 0.152$ & 0.0045 \\
        $I_c - i = -0.409\,(i-z) - 0.394$ & 0.0063 \\
    \hline
    \hline
        \textbf{Relation to go from USNO-B1 to Johnson-Cousins (\citealt{usnotransfo})} & \textbf{Error} \\
    \hline
        $B = B1$ & 0.5 \\
        $R_c = R1$ & 0.5 \\
        $V = 0.444\,B1 + 0.556\,R1$ & 0.5 \\
    \hline
    \hline
        \textbf{Relation to go from Pan-STARRS to Johnson-Cousins (\citealt{PStransfo})} & \textbf{Error} \\
    \hline
        $B - g = 0.016\,(g-r)^2 + 0.540\,(g-r) + 0.199$ & 0.056 \\
        $V - g = -0.008\,(g-r)^2 - 0.498\,(g-r) - 0.020$ & 0.032 \\
        $R_c - r = -0.061\,(g-r)^2 - 0.086\,(g-r) - 0.163$ & 0.041 \\
        $I_c - i = -0.263\,(g-r)^2 - 0.040\,(g-r) - 0.433$ & 0.048 \\
    \hline  
    \enddata
\end{deluxetable}

\subsection{Estimation and Subtraction of the Background Level}
\label{bkg}
The evaluation of the background level in the image is performed with the \texttt{Photutils} library \citep{phot}. The algorithm meshes the image in $N$ square boxes $Bo$, whose size is chosen by the user such that it is small enough to capture the local background variations and larger than the typical size of sources in the image. In each box, the background is estimated with a user-chosen method. By default, we use the same estimator as \texttt{SExtractor} \citep{sex}: 
$$
{\rm Background} = 
\left\{
    \begin{array}{ll}
       2.5 \, \, \tilde{Bo}_{i}^{N} - 1.5 \,\, \hat{{Bo}}_{i}^{N} & \mbox{, if } \frac{(\hat{Bo}_{i}^{N} - \tilde{Bo}_{i}^{N})}{ \sigma_{Bo_{i}^{N}}} < 0.3 \\\\
     \tilde{Bo}_{i}^{N}  & \mbox{, otherwise,}
    \end{array}
\right.
$$ 
where $\tilde{Bo}_{i}^{N}$ and  $\hat{Bo}_{i}^{N}$ are respectively the median and mean pixels values in $Bo$.
This step is illustrated in Figure~\ref{images}, where the two-dimensional background images are shown in the middle column, and the background-subtracted images are in the right-hand column.

\subsection{Source Detection and Photometry}
\label{photo}
In the background-subtracted image, \pip~ considers a source to be detected if a user-chosen number of connected pixels are twice the standard deviation above the background. The choice of connected pixels must be done according to the pixel size of the image and the typical size of the objects: a large number of pixels will lead to missing the faintest objects. On the other hand, a low number of pixels will end up with an image polluted with hot pixels and cosmic rays. A default value of 5 gave a good compromise for the tests we performed.

Using the first-order moment, we compute the sources' centroids both in physical and celestial coordinates. Their corresponding fluxes are estimated using one of the three following approaches (chosen by the user).
\begin{itemize}
    \item \textbf{Isophotal Aperture\footnote{\url{https://photutils.readthedocs.io/en/stable/segmentation.html}}:} The fluxes are estimated in elliptic apertures whose semimajor and semiminor axes lengths and orientations are computed using the second-order moment of the sources.
    The axial lengths determined this way are equivalent to spatial standard deviations computed in the axial direction. Thus, to avoid underestimating the source flux in the aperture, the two axes lengths are multiplied by a coefficient (the default is 3) called the isophotal coefficient. 
    \item \textbf{Fixed Aperture\footnote{\url{https://photutils.readthedocs.io/en/stable/aperture.html}}:} The fluxes are estimated in circular apertures centered on the centroids of the detected sources. The extension of the apertures is fixed for all the sources and computed by multiplying the image mean full width at half-maximum intensity (FWHM) of the PSF estimated with \texttt{PSFex} by a fixed factor that is user-selected (the default is 3). Two criteria must guide the choice of its value: having a large enough aperture to avoid flux underestimation, but not too large to avoid contamination by surrounding sources.
    \item \textbf{Kron Aperture:} The fluxes are determined in ellipses whose parameters are estimated following the method described by \citep{kron}. The ellipses parameters (orientation, semiminor axis, semimajor axis) are estimated with the second-order moments. The extension of the apertures is evaluated with the first-order moments of the source.
\end{itemize}

For each algorithm, we use the number of counts in analog-to-digital units (ADU) in the aperture to compute the instrumental magnitude:
\begin{equation}
    m_{\rm ins} = -2.5\,  {\rm log}_{10} \left(N_{\rm ADU} \right).
    \label{eq_mag}
\end{equation}

\subsection{Photometric Calibration}
\label{calibration}
The catalog we chose at the step described in Section~\ref{cata} is used for crossmatching the sources detected using their centroids. Each source is associated with the closest object in a radius of $5''$ in the catalog as visible in the left column of Figure~\ref{images}. If available, we use data-quality flags given by the survey to crossmatch only stellar objects with good photometry. Table~\ref{tab:flag} presents the flag used for the four available catalogs. For images acquired in a filter that differs from the catalog bands, we use the transformation relations defined in Table~\ref{tab:filter} to change from the survey's photometric system to the image's band.

\begin{figure}[htbp]
  \centering
  \includegraphics[width=1.0\textwidth]{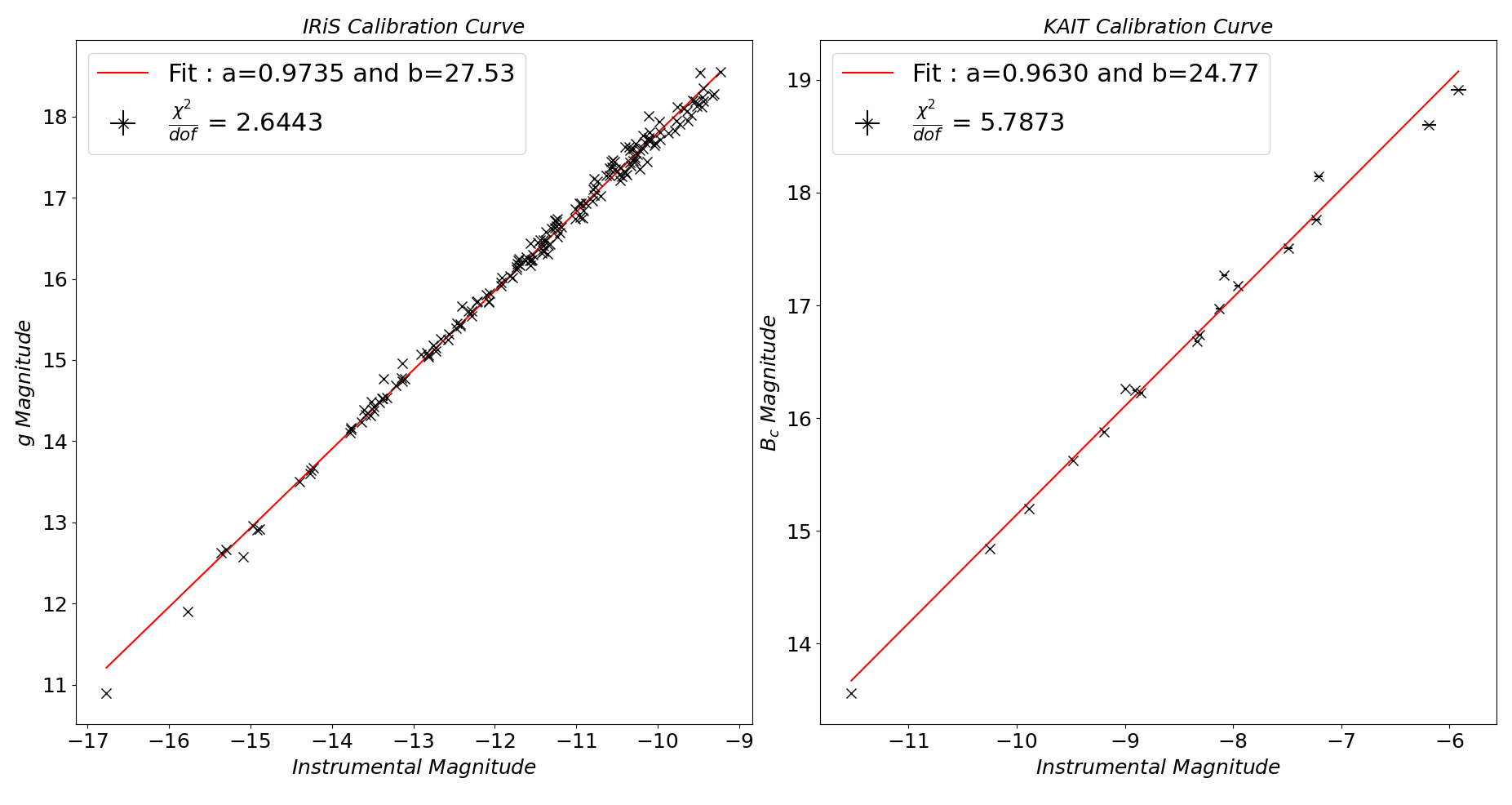}
  \caption{Calibration relations we obtained for IRiS and KAIT. The black crosses are the sources detected in the image and crossmatched with the catalog. The calibration of the sources of interest, including the monitored objects and the star used to evaluate the image quality, is done with the linear fits shown in red.}
  \label{calib_curve}
\end{figure}

The final photometric calibration of an image is done by fitting a linear relation between the instrumental magnitude of the detected objects and their calibrated magnitude. The fits are plotted in red in Figure~\ref{calib_curve} for two example images from IRiS on the left and KAIT on the right. This fit is used to compute the calibrated magnitude of the sources of interest in the image, which includes the transient and the star chosen as reference for the quality check described in Section~\ref{veto} The precision of this method depends mainly on the number of objects visible in the field of view --- the more detected objects, the better the precision. This is detailed in Sections~\ref{error_dis} and \ref{error_prop}). 

The systematic uncertainties tied to this method are estimated to be 0.05--0.1\,mag for IRiS images that have several tens of sources in the field of view, and up to 0.3\,mag for fields of view with only $\sim10$ stars.
The use of catalogs with a filter different from the one with which the image was acquired will also increase the uncertainty, as can be seen in Table~\ref{tab:filter}), typically 0.05\,mag for SDSS, Pan-STARRS, and {\it Gaia}. For the USNO-B1 catalog, the uncertainties are typically 0.5\,mag. These large uncertainties are due to the heterogeneity of the data and instruments used for photometry.

We caution that the results provided by this method are given in the natural system, not in a standard system. Consequently, if there is a significant discrepancy between the instrument's filter passband and the standard filters from SDSS, Pan-STARRS, or the Landolt system \citep{landolt1,landolt2}, an additional color term may be needed.

\begin{deluxetable}{ccc}
\label{tab:flag}
    \tablewidth{0.72\textwidth}
    \tablecaption{Quality flags used for the crossmatch with a reference catalog.}
    \tablehead{ \colhead{Survey} & \colhead{Keyword}  & \colhead{Definition}  }
    \startdata
        \hline
                    & Qual (Binary Flags) & Classified extended by Pan-STARRS \\
                    & Qual (Binary Flags) & Good measurement in PS \\
        \href{https://vizier.u-strasbg.fr/viz-bin/VizieR-3?-source=II/349}{Pan-STARRS}
                    & Qual (Binary Flags) & Only one good stack measuremement \\
                    & Qual (Binary Flags) & suspect object in the stack \\
                    & Qual (Binary Flags) & Poor-quality stack object\\
        \hline
                & q\_mode & Quality of the photometry \\
                & Q       & Bad Sky flag \\
                & class   & Type of object (star or galaxy)\\
        \href{https://vizier.u-strasbg.fr/viz-bin/VizieR-3?-source=V/147}{SDSS}
                & flag (hexadecimal flag) & Saturated Pixel \\
                & flag (hexadecimal flag) & Object on the Edge \\
                & flag (hexadecimal flag) & Blended Object \\
                & flag (hexadecimal flag) & Moving Object \\
        \hline
        \href{https://vizier.u-strasbg.fr/viz-bin/VizieR-3?-source=I/345/gaia2}{Gaia}
        & Dup & Duplicated source$^a$ \\
        \hline
        \href{https://vizier.u-strasbg.fr/viz-bin/VizieR-3?-source=I/284/}{USNO} & Ndet & Number of detection $\leq$ 1 \\
                                  & Flags & Object on a diffraction spike \\
        \hline
    \enddata
    \tablecomments{$^a$ This may indicate observational, crossmatching or processing problems, or stellar multiplicity, and probable astrometric or photometric problems in all cases.}
\end{deluxetable}

\subsection{Error Propagation}
\label{error_prop}
Several effects are taken into account to quantify Muphoten's uncertainties, as follows. 
\begin{itemize}
    \item \textbf{Poisson noise:} The number of ADU associated with the transient has an uncertainty given by the Poisson distribution, $N_{\rm \mathbf{ADU}} \pm \sqrt{N_{\rm \textbf{ADU}}}$. For the instrumental magnitude, the uncertainty is then
    $$\delta m_{\rm ins} = \frac{2.5}{{\rm ln}(10)\sqrt{N_{\rm ADU}}}.$$
    
    \item \textbf{Background level around the transient: }The number of ADU from the background, $N_{\rm background}$, is computed using the source of interest's aperture on the background image (see the central column of Figure~\ref{images}). The uncertainty in the calibrated magnitude from the background is given by
    $$\sigma_{\rm background} = \frac{2.5}{{\rm ln}(10)\sqrt{N_{\rm background}}}.$$ 
    
    \item \textbf{Calibration error: }To compute the calibrated magnitude from the instrumental magnitude, we use a linear fit as described in Section~\ref{calibration}), $m_{\rm calibrated} = a\,m_{\rm ins} + b$. There are uncertainties in the parameters $a$ and $b$ ($\delta a$ and $\delta b$) given by the fit and in the instrumental magnitude ($\delta m_{\rm ins}$) given by the Poisson distribution. These uncertainties are propagated as
    $$\sigma_{\rm calibration} = \sqrt{m_{\rm ins}^{2} \delta a^2 + \delta b^2 + a^2 \delta m_{\rm ins}^2}.$$

    \item \textbf{Transformation filter: }The transformation relations used to go from one photometric system to another introduce an uncertainty, $\sigma_{\rm transfo}$, whose value depends on the method used. They are given by \cite{gaiatransfo} for {\it Gaia}, \cite{usnotransfo} for USNO-B1, \cite{sdsstransfo_paper} for SDSS, and \cite{PStransfo} for Pan-STARRS, and are summarized in Table~\ref{tab:filter}.

\end{itemize}

These sources of uncertainty are independent and thus are added in quadrature to compute the total uncertainty, 
\begin{equation}
\label{eq_error}
\sigma_{\rm tot} = \sqrt{\delta m_{\rm ins}^2 + \sigma_{\rm background}^2 + \sigma_{\rm calibration}^2 + \sigma_{\rm transfo}^2}.
\end{equation}

There is no aperture correction taken into account in \pip~. For Kron photometry, the method that computes the aperture recovers at least 90\% of the flux of the source according to \citet{kron}. For the two other options available in this work, the values for the coefficients used to scale the apertures must be chosen to be large enough to avoid the underestimation of the sources fluxes, but small enough to avoid contamination by other sources. For more details about the aperture correction, see Section 2.5.1 of \citet{these_pipien}.
The background level is subtracted before conducting photometry to avoid including it in the source flux estimation, which would degrade the magnitude estimate.

\subsection{Additional Features of the Pipeline}
\subsubsection{Quality Checks}
\label{veto}
Follow-up observations of a transient source can be performed over long timescales (several weeks to months) with potentially many independent telescopes. This may lead to significant variations in quality among the images. For this reason, having data-quality checks to reject images with bad observing conditions is necessary. Consequently, we implemented two automated quality checks for this work: the first is based on the PSF, and the second is based on the temporal stability of the photometry of a known, nonvariable star within the field of view.

Using systematic vetoes for follow-up images of a transient constitutes one original aspect of \pip~ compared to what is usually done when several heterogeneous telescopes observe a transient; see \citet{kilo17} and \citet{follow17} for GW\,170817, or \citet{cowprentice}, \citet{cowperley}, and \citet{cowmarg} for SN\,2018cow.

\begin{itemize}
    \item \textbf{PSF check: }
    We use the open-source software \texttt{PSFex} \cite{psfex} to compute the mean FWHM of the PSF over the frame. Then, for the images from the same telescopes and in the same band, we build the distribution of the PSF's FWHM mean and remove the images that deviate from the median of this distribution by more than 3$\sigma$. This procedure ensures that the large datasets are homogeneous by rejecting images with data-quality issues (see Figure~\ref{psf}).

    \begin{figure}[htbp]
        \centering
        \includegraphics[width=0.70\textwidth]{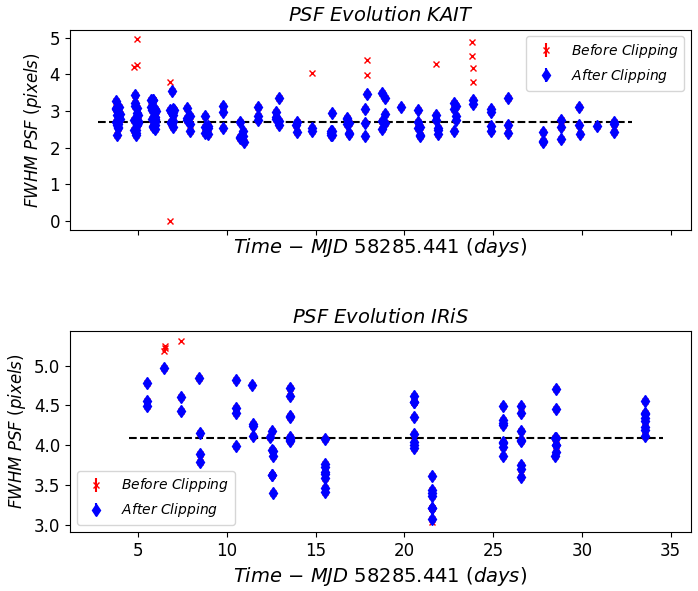}
        \caption{Evolution of the PSF's FWHM with time for the images of SN\,2018cow taken by KAIT (upper plot) and IRiS (lower plot). The black dashed lines are the median of the PSF's FWHM distributions, the blue diamonds are the images with a mean FHWM's PSF over the frame that deviate by $< 3\sigma$ from the median, and the red crosses are for the images that are above that limit and then rejected from the dataset via sigma-clipping.}
         \label{psf}
    \end{figure}

    \item \textbf{Light curve of a reference nonvariable star:}
    We compute the light curve of a user-selected star in the field of view with the same photometric calibration as the transient to test the image quality. The star should be visible in all images of the dataset. If the computed magnitude and the catalog magnitude are incompatible (e.g., the intervals given by the respective uncertainties are not overlapping) or if the total uncertainties are larger than a user-selected threshold on the error bars, the image is rejected.
    Figure~\ref{star} shows the light curve of a star chosen to test the calibration of the images produced during the monitoring of SN\,2018cow. Figure~\ref{ex_veto} show an example of an image rejected by this quality check.
    The threshold is telescope-dependent in order to account for characteristics such as location or field of view. For example, we use a 0.15\,mag threshold for KAIT, but a 0.10\,mag for IRiS as it has a larger field of view and thus more stars for calibration. Consequently, this threshold needs to be tuned in practice for every instrument. This check helps to evaluate both the image quality and the photometric calibration.
    
    \begin{figure}[htbp]
        \centering
        \includegraphics[width=0.80\textwidth]{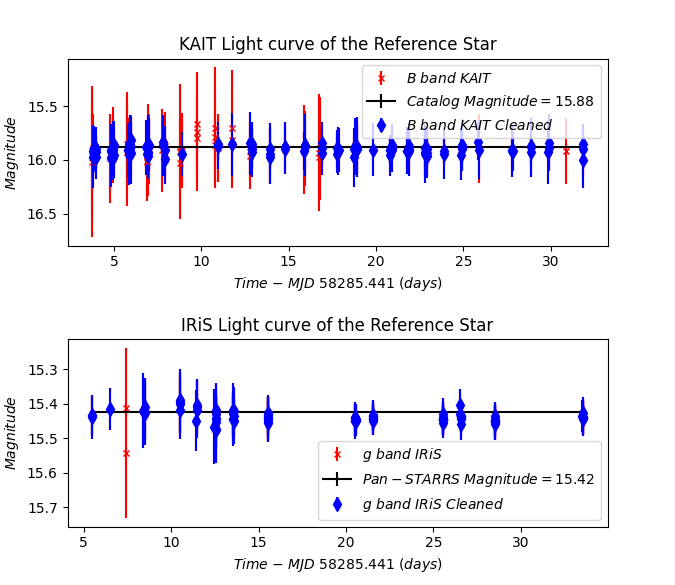}
        \caption{Light curves of a star located at $\alpha=243.97494^\circ$, $\delta=22.29366^\circ$.
                The upper plot is the light curve of the star used for evaluating the image quality for KAIT in the $B$ band, and the bottom plot corresponds to the light curve for IRiS in the $g$ band.
                Both were calibrated using Pan-STARRS stars and are given in the AB system 
               \citep{1983ApJ...266..713O}. Blue diamonds correspond to the images passing the check, and the red crosses correspond to the rejected images.} 
         \label{star}
    \end{figure}

    \begin{figure}[htbp]
        \centering
        \includegraphics[width=1.0\textwidth]{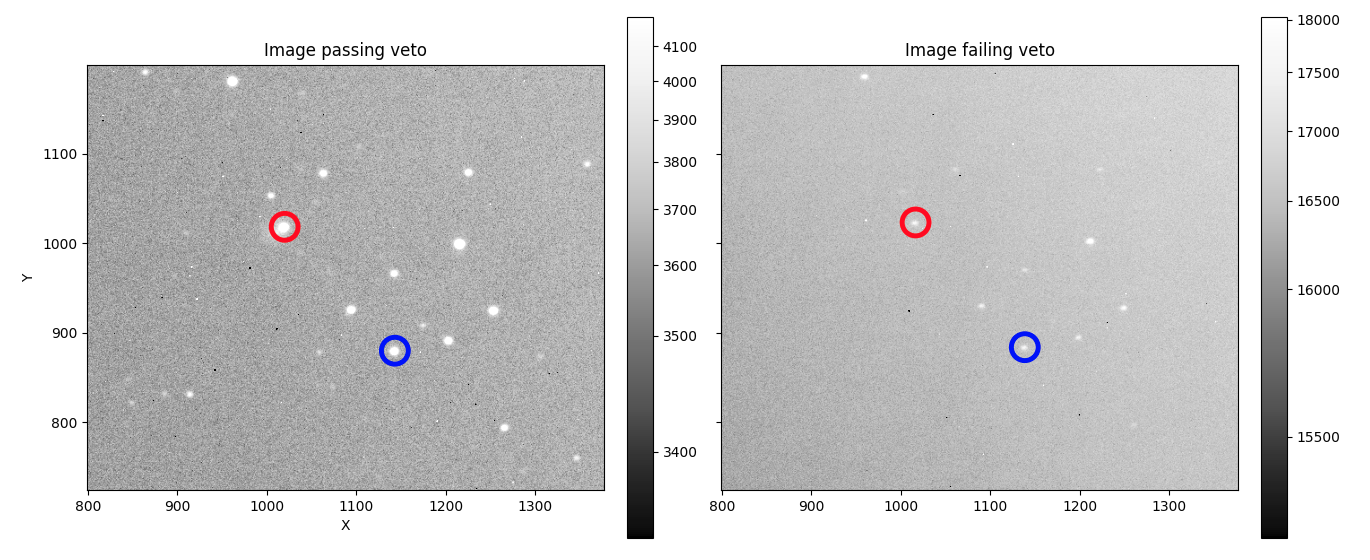}
        \caption{Example of an image rejected by the reference star quality check.
                The left panel shows an image acquired by IRiS on its first night of observation that passes the veto. The right panel shows an image that has been rejected owing to a particularly bright background sky. The red circles correspond to the AT\,2018cow transient, and the blue circles are for the star used as a veto}. The rejected image is visible as one of the red points at $T \approx 7$ days in the bottom plot of Figure~\ref{star}.
         \label{ex_veto}
    \end{figure}

\end{itemize}

\subsubsection{Extracting the Limiting Magnitude of an Image}
\label{section_mag_lim}
\begin{figure}[htbp]
      \centering
      \includegraphics[width=1.0\textwidth]{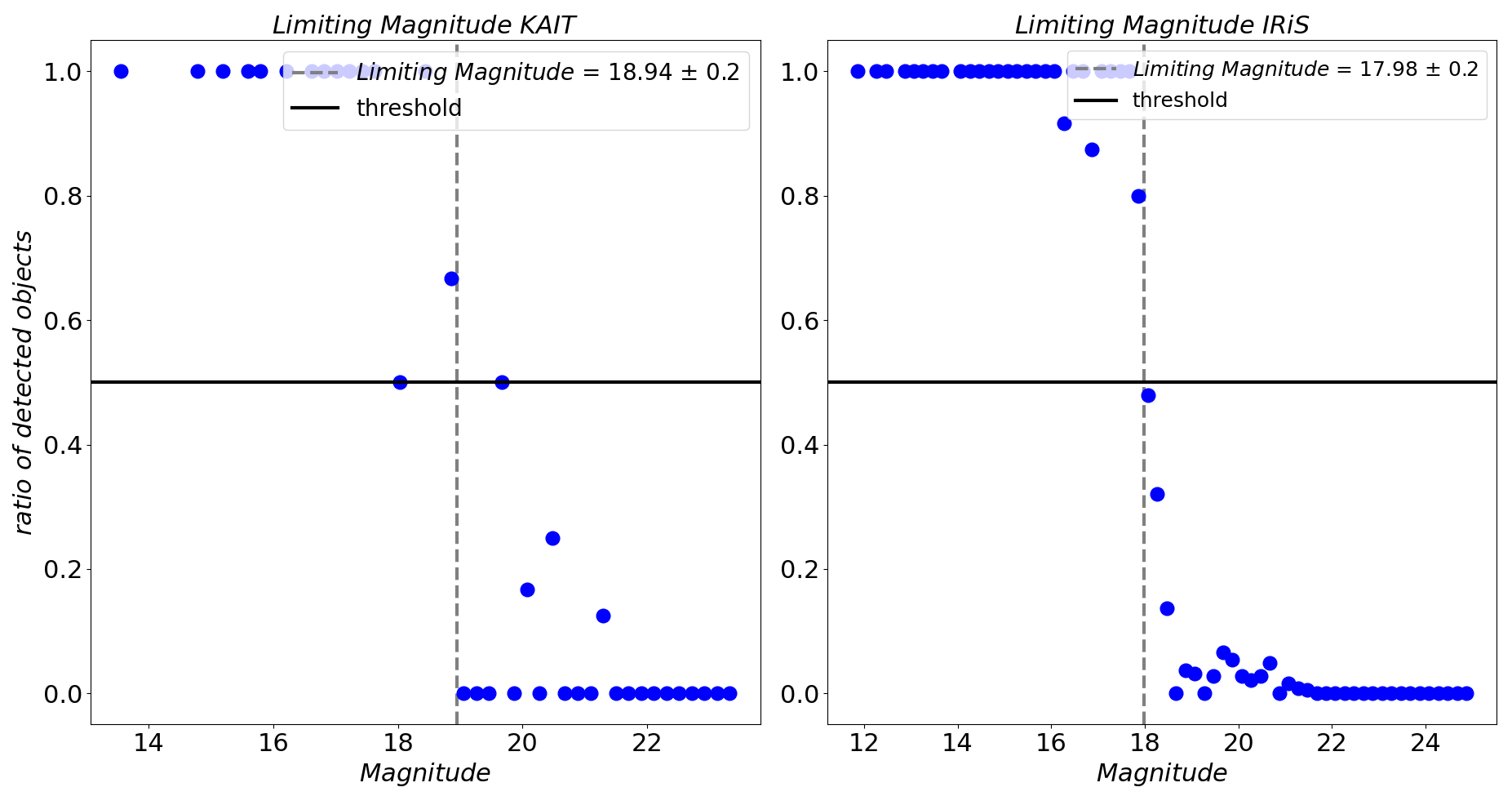}
      \caption{Limiting magnitude for KAIT (left) and IRiS (right).
      Blue dots are the ratio of detected sources divided by the number of sources in Pan-STARRS in bins of width 0.2\,mag. The gray dashed line is for the limiting magnitude, and the black horizontal line is the threshold to consider that the limiting magnitude is reached, here 50\%.}
      \label{mag_lim}
\end{figure}

The limiting magnitude is estimated by computing the ratio of detected objects in the image and the number of objects in a reference catalog in magnitude bins. When the ratio drops below 0.5 and if it does not rise above this threshold for larger magnitudes, the center of the corresponding bin is considered to be the limiting magnitude, as seen in Figure~\ref{mag_lim}). The user selects the width of the bin which determines the precision of the limiting-magnitude estimate.

Although this method relies on an external catalog to estimate the limiting magnitude, which can limit the depth one can use, it has several advantages for a network such as GRANDMA. First, small-aperture instruments that constitute most of GRANDMA and its amateur-branch telescopes will not have deeper images than Pan-STARRS. Consequently, this method is not limiting for that case. 
In addition, low-latency observations of transient events require the ability to rapidly distribute information to the community via GCN, for example. It includes the upper limits of the observation to preclude useless observations of objects too faint for small instruments. The method presented here can give rapid image-depth estimates for numerous frames acquired by heterogeneous instruments. These estimates can efficiently be used for follow-up observations of gravitational-wave or gamma-ray burst alerts. However, another method must be used for cases where the upper limits have to be precisely evaluated or if the image is deeper than the Pan-STARRS limiting magnitude. Currently, efforts are made to implement other limiting magnitude estimates in \pip~. The method we are currently testing starts by measuring the background scatter around the target. Then, we measure a few other places sampled in the images and take their average as the average background. We then use three times the average scatter to calculate the 3$\sigma$ limiting magnitude. On the other hand, we are also putting some effort into implementing a method for injecting fake stars into the frames to have a precise estimate. However, this method is computationally expensive and is planned to be used for the deepest images.

\section{\textbf{Application}}
\label{application}
The method presented here has been tested on a transient detected in 2018: SN\,2018cow/AT\,2018cow (or simply ``the Cow"; \citealt{ATGBM}; \citealt{milli17}; \citealt{ATLAT}; \citealt{cowswift}; \citealt{cowmarg}; \citealt{hotecow}; \citealt{cowperley}; \citealt{ATsmart}). This transient belongs to an emerging class of optical transients called luminous fast-rising blue optical transients (LFBOTs; see Section~\ref{lavache}). In this section, we present the results of the \pip~analysis of the Cow images acquired by five different telescopes. Muphoten's results are then compared to the light curves published elsewhere \citep{cowswift,cowmarg,cowperley,cowprentice}. The images we used were taken by IRiS, KAIT \citep{kaitpap}, TAROT Chile (TCH; \citealt{tarot}), the Kitt Peak EMCCD demonstrator (KPED) on the Kitt Peak 2.1\,m telescope \citep{kped}, and the Liverpool Telescope (LT; \citealt{lt}).
Using the images from the LT and KPED for which the light curves are already published helped us to validate our method by comparing our results to an independent analysis. We also used unpublished images from IRiS, KAIT, and TCH that returned results compatible with published light curves.

\subsection{\textbf{SN\,2018cow}}
\label{lavache}
\subsubsection{Discovery and Properties}
The Cow was detected by ATLAS, a 0.5\,m telescope, on 2018 Jun 16 10:35:38 UTC (which corresponds to MJD 58285.441) at $o = 14.74 \pm 0.1 $\,mag (orange $o$ band) and no previous detection on MJD = 58281.5 up to $o \approx 19.5$\,mag, indicating an unusually fast rise time. Its location, $\alpha$(J2000) $= 16^{\rm h}16^{\rm m}00.22^{\rm s}$ and $\delta$(J2000) $= +22^\circ 16' 04.8''$, is coincident with the dwarf starburst galaxy CGCG 137-068 located at a distance of $66 \pm 5$\,Mpc and with an offset of 1.7\,kpc from the galaxy's center.

At first, the Cow was announced as a cataclysmic variable (CV) by \citep{ATsmart}. After 2.6\, days, an LT spectrum \citep{cowprentice,cowperley} revealed a featureless, hot, blue transient, ruling out the CV hypothesis. The Cow has been detected across the entire electromagnetic spectrum, except in gamma rays \citep{ATGBM,ATLAT}. The X-ray light curve shows several episodes of rebrightening, most probably burst-type events rather than periodic activity \citep{cowswift}, suggesting a central engine \citep{cowmarg}. The millimetric \citep{milli17} and radio \citep{hotecow} counterparts of SN\,2018cow were detected at early times and are relatively bright. 

\subsubsection{Observations and Muphoten Configuration}
\label{result_vache}
Observations by IRiS in the $g$ band started on 2018 June 21 and lasted until Sep. 2018. The subtraction was performed with a template image taken on 2018 Sep. 07. KAIT observed the Cow in the $B$ band from 2018 June 20 until 2018 Aug. 08. The template used for subtraction was acquired on 2018 Aug. 07. The LT conducted its observations from 2018 June 19 until 2018 Oct. 01 in the $g$ band and from 2018 June 20 to 2018 Sep. 30 in the $B$ band. For the LT images in the $g$ band, we used a Pan-STARRS frame as the template image, and for the $B$ ones, we used the last image available in the LT $B$-band dataset as a template. KPED observed the Cow from 2018 June 20 until 2018 July 07 in the $g$ band, and the template image was a Pan-STARRS frame. TCH observed from 2018 June 21 until 2018 June 24 in $g$, and a Pan-STARRS image was used for subtraction. We chose to use templates constructed with Pan-STARRS frames and from the same telescope archive to test the two cases possible in {\tt Muphoten}.

We evaluated the background with the \texttt{Sextractor} estimator and calibrated the images against Pan-STARRS stars for all five telescopes. We used isophotal photometry with an extension of 5 for all images. For calibrating KAIT and the LT $B$ images, we used the transformation relations in Table~\ref{tab:filter} to go from Pan-STARRS's photometric system to the image's band. 

The telescopes we used for this work have their characteristics summarized in Table~\ref{tab:telescopes}. The field of view, diameter, and sampling spanning values over an order of magnitude constitute a significant difficulty for the reduction of images acquired by a heterogeneous network of telescopes.

\begin{deluxetable}{cccccc}[htbp]
    \label{tab:telescopes}
    \tablewidth{0.72\textwidth}
    \tablecaption{Summary of the instruments that monitored SN\,2018cow.}
    \tablehead{ \colhead{Telescope} & \colhead{Diameter}  & \colhead{Focal Ratio} & \colhead{Field of View} & \colhead{Filter} & \colhead{Sampling} \\ 
                \colhead{ } & \colhead{[m]}  & \colhead{ } & \colhead{} & \colhead{ } & \colhead{[arcsec/pixel]}  }
    \startdata
        KAIT      & 0.76     &  $f/8.2$  &  $6.67' \times 6.67'$  & Clear, $UBVR_c I_c$ &  0.8          \\ 
        TCH       & 0.25     &  $f/3.2$  & $1.8^\circ \times 1.8^\circ$ & Clear, $griz$ &  3.2    \\ 
        IRiS      & 0.50     & $f/8.2$ &      $24'$       &  $griz$        &     0.7       \\ 
        LT        & 2.0      &  $f/3$  &  $10' \times 10'$       &  $BV$, $ugriz$   &     0.15      \\ 
        KPED      & 2.1      & $f/4.86$ &   $4.4' \times 4.4'$   & $UBVR_c I_c$, $gr$    &    0.26       \\ 
    \enddata
\end{deluxetable}


\subsubsection{Vetoes}
We summarize the numbers of images we analyzed to obtain Figure~\ref{lccow} in Table~\ref{tab:veto}. For KPED and TCH, we had images of the Cow when it was still very bright and clearly detected in all the images. For the other instruments, KAIT had the most images (68) in which the transient was not detected owing to its limiting magnitude ($\sim 18$\,mag). The same explanation stands for IRiS that monitored the Cow for two months with a limit of $\sim 18$\,mag, leading to no detection of the Cow after 35\,days.

For the veto based on the reference-star light curve, between 1 and 5 images were rejected for TCH, KPED, IRiS, and LT. For KAIT, 50 images were rejected, indicating calibration difficulties because of the small number of sufficiently bright sources in the field of view. For TCH, except for two images in which the star had a magnitude incompatible with the catalog value, the rejection occurs because of larger photometric uncertainties. These thresholds were 0.25\,mag for TCH, 0.8\,mag for KEPD, 0.3\,mag for KAIT, and 0.15\,mag for IRiS and the LT; see Section~\ref{error_dis} for a discussion of these values.

The PSF-based veto did not reject any images for TCH and KPED. Between 1 and 6 images were rejected for IRiS, the LT, and KAIT. After visual inspection, these images had very poor seeing and thus were impossible to analyze.

\begin{deluxetable}{cccccc}[htbp]
    \label{tab:veto}
    \tablewidth{0.72\textwidth}
    \tablecaption{Summary of the number of images processed with {\tt Muphoten}.$^a$}
    \tablehead{ \colhead{Telescope} & \colhead{Images Processed}  & \colhead{Nondetection} & \colhead{Rejected Star Veto} & \colhead{Rejected PSF Veto} & \colhead{Images Remaining}}
    \startdata
        TCH     	 & 17  & 0  & 4  & 0 & 13 \\
        KPED     	 & 39  & 0  & 5  & 0 & 34 \\
        IRiS     	 & 111 & 17 & 2  & 2 & 90 \\
        LT: $g$      & 161 & 42 & 1  & 6 & 112\\
        LT: $B$      & 63  & 17 & 1  & 1 & 44 \\
        KAIT 		 & 239 & 68 & 50 & 6 & 115\\
    \enddata
    $^a${We indicate the number of frames where the transient is detected along with the number of images rejected by the two vetoes.}
\end{deluxetable}

\subsubsection{Error Discussion}
\label{error_dis}
As described in Section~\ref{error_prop}, we consider three sources for the uncertainties in the pipeline: Poisson, calibration, and background errors. Except for KPED, the dominant uncertainty is the calibration, as the signal-to-noise ratio is sufficiently high to make both Poisson and background uncertainties negligible. The lower Poisson and background uncertainties of IRiS compared to the LT are due to the longer exposure time of the first (300\,s) compared to the second (60\,s). KAIT calibration errors are $\sim 0.2$\,mag because of the few detected objects in its field of view (typically $\sim 10$ sources; see Figure~\ref{calib_curve}). This leads to higher uncertainties in the fit parameters and, therefore, higher uncertainties in the calibration. The latter argument also explains why the calibration errors are larger for the LT than for IRiS, as the first detects fewer sources in its smaller field of view. For KPED, the much larger error bars are due to a short exposure time (10\,s) that leads to a low ADU count and consequently to large Poisson and background error bars (see Eq.~\ref{eq_error}). Moreover, the lower number of detected sources ($\sim 20$) increases the calibration errors. 

\begin{deluxetable}{cccc}[htbp]
    \label{tab:Error}
    \tablewidth{0.99\textwidth}
    \tablecaption{\textbf{Summary of the uncertainties for each telescope.$^a$}}
    \tablehead{ \colhead{Telescope} & \colhead{Poisson Error}  & \colhead{Calibration Error} & \colhead{Background Error} \\ 
                \colhead{-} & \colhead{[mag]}  & \colhead{[mag]} & \colhead{[mag]} }
    \startdata
        TCH     	 & 0.009 & 0.12 & 0.002  \\
        KPED     	 & 0.31  & 0.32 & 0.52  \\
        IRiS     	 & 0.008 & 0.06 & 0.002   \\
        LT: $g$       & 0.01  & 0.08 & 0.006   \\
        LT: $B$       & 0.02  & 0.08 & 0.01   \\
        KAIT 		 & 0.03 & 0.2 & 0.01  \\
    \enddata
    $^a${The mean errors are computed for the images that passed the two vetoes described in Section~\ref{veto}.}
\end{deluxetable}

\subsubsection{Results}
\begin{figure}[htbp]
  \centering
  \includegraphics[width=0.75\textwidth]{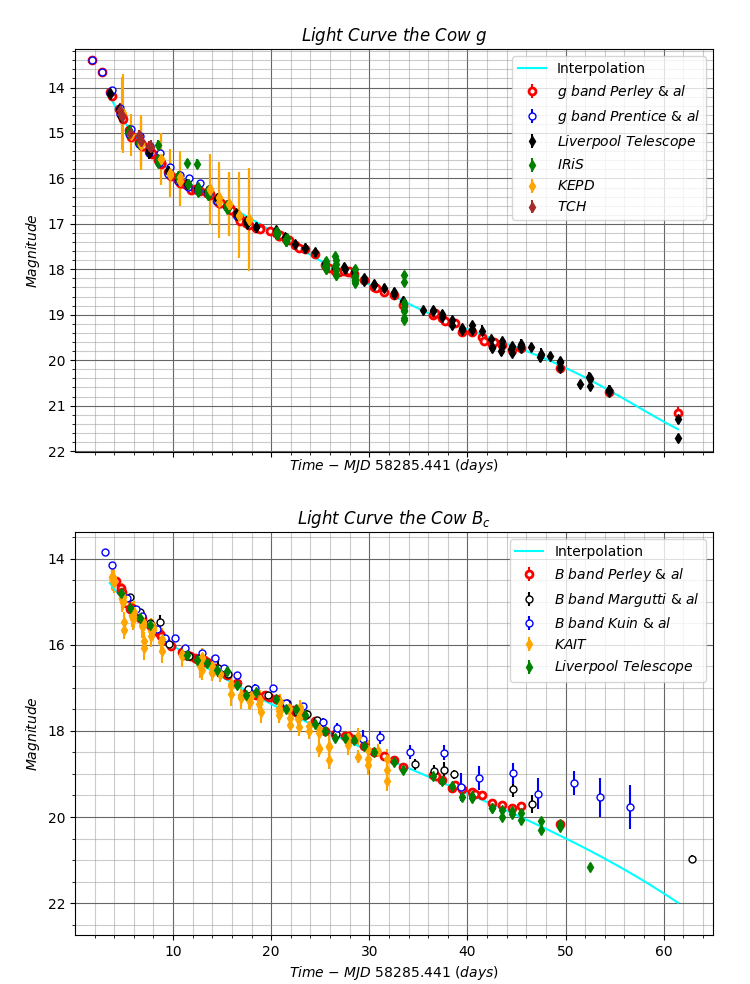}
  \caption{Light curve for the Cow in the $g$ (top) and $B$ bands (bottom) in the AB system.
          The circles are the points extracted from the literature about the Cow (\citealt{cowperley}; \citealt{cowmarg}; \citealt{cowswift}; \citealt{cowprentice}). The diamonds are from this work. In the $g$ band, for the last night IRiS detected the transient, the measurements show a large dispersion around $\sim 18.5$\,mag}. As these images passed our vetoes, this is not due to quality nor a calibration issue. \textbf{Visual inspection of those images and the calibration curves does not reveal any issues. Consequently, we attribute} this dispersion to the \textbf{template subtraction becoming more sensitive to small image variations as the transient brightness approaches the} IRiS limiting magnitude. The faintness makes the measurements more sensitive to small variations in subtraction among images in the same night.
  \label{lccow}
\end{figure}

The final results are presented in Figure \ref{lccow} for the $B$ and $g$ bands along with the results published by \citet{cowperley}, \citet{cowmarg}, \citet{cowswift}, and \citet{cowprentice}. Our results are in very good agreement with those in the literature. They demonstrate that  \pip~ can produce homogeneous datasets of an astrophysical transient observed by heterogeneous instruments with various fields of view, locations, bands, and sensitivities. This photometry is also consistent with measurements performed by independent teams with different instruments and by the LT and KPED used by \citet{cowperley}.

\begin{figure}[htbp]
  \centering
  \includegraphics[width=0.70\textwidth]{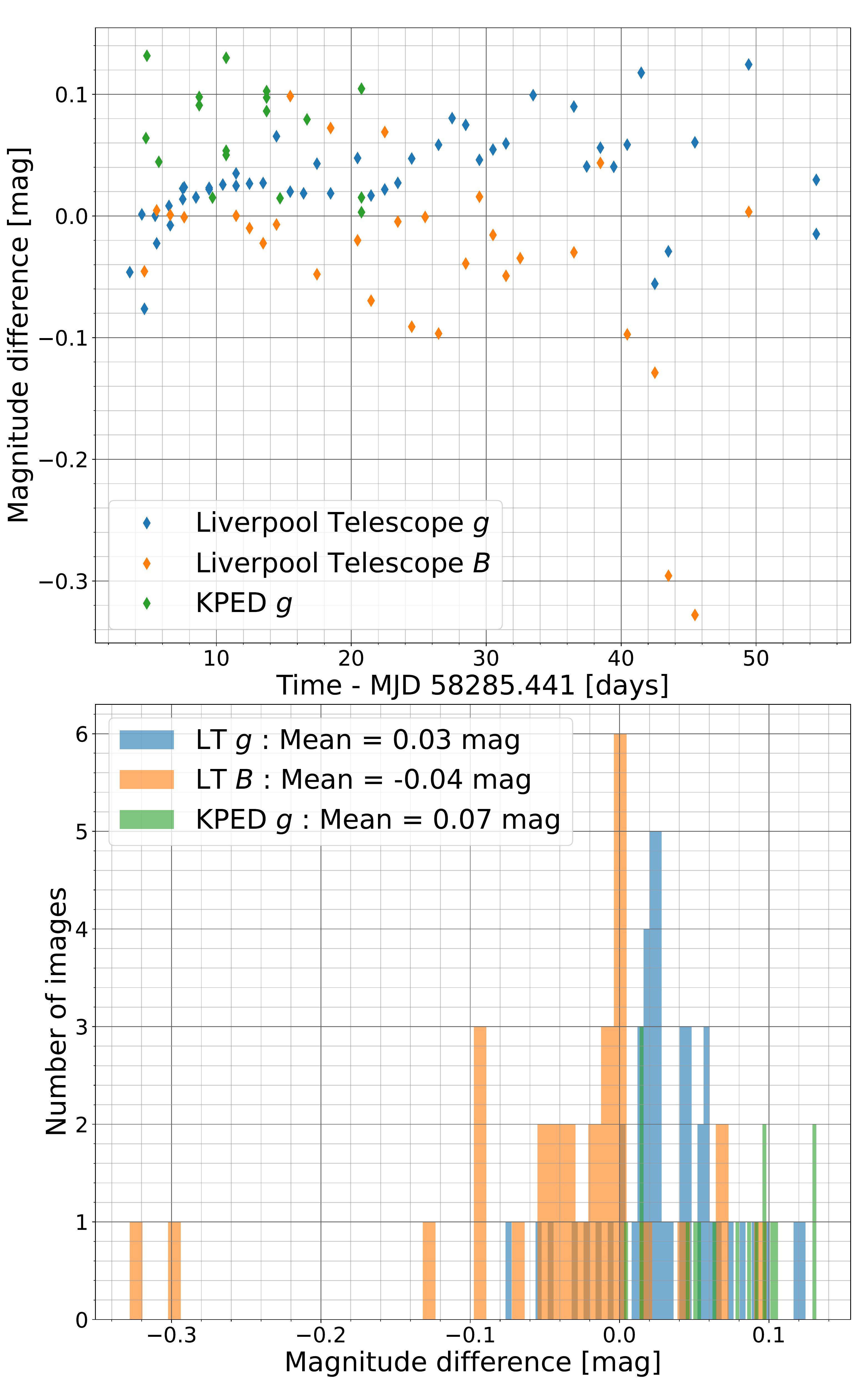}
  \caption{Magnitude differences between the \pip~ results and the results of \citet{cowperley} for the LT and KPED images used in both studies.
           The upper plot corresponds to the temporal evolution of the magnitude difference for the $g$ and $B$ bands of the LT (in blue and orange, respectively) and the KPED in green. The bottom plot corresponds to the magnitude differences for the three cases: $g$ of LT in blue, $B$ of LT in orange, and $g$ of KPED in green.}
  \label{magdiff}
\end{figure}

To evaluate \pip~ results, we compared the measurements for the images used by both \citet{cowperley} and this work. They include images by the LT in both $g$ and $B$, and $g$-band images by the KPED instrument. On the one hand, we evaluated the temporal evolution of the difference between the {\tt Muphoten} and Perley et al. measurements. The upper plot of Figure~\ref{magdiff} shows the results for LT and KPED. For the LT, the results are consistent within $\pm 0.1$\,mag, except for a few points beyond 40\,days. Moreover, past this delay, the dispersion increases in both bands. All these images pass the quality checks implemented in \pip~, and none show any issue after visual inspection. Consequently, we attribute this dispersion to the transient becoming fainter, which makes the template subtraction with {\tt HOTPANTS} more sensitive to differences between the configuration used for this work and the literature. The subtraction sensitivity for transients becoming too faint for a telescope is also visible in Figure~\ref{lccow}. The last night that IRiS detects the transient, the dispersion among the images acquired during the night increases, despite no visible issue in the images. For the KPED images, the results show larger dispersion compared to the $g$ band of the LT at the same time. However, Figure~\ref{lccow} shows that the uncertainties in the magnitudes for the KPED images are larger than for the LT, which makes those results consistent with the literature.

On the other hand, we also plotted the magnitude differences as histograms in the bottom panel of Figure~\ref{magdiff}. This demonstrates that the differences between \pip~ results and the literature are $< 0.1$\,mag. Considering the typical uncertainties expected to be 0.1\,mag for the LT and 0.5\,mag for KPED (according to Table~\ref{tab:Error}), the results we obtain with \pip~ are considered to be consistent with the literature. The few outliers in these histograms correspond to LT images acquired more than 40\,days after the Cow's detection, which correspond to images where subtraction becomes more difficult. In addition, we evaluated the mean difference between the two LT bands and the KPED datasets. For the $g$ band, we found 0.03\,mag and 0.07\,mag for the LT and KPED, respectively, and $-0.04$ for the LT $B$ band. These values are lower to the typical magnitude uncertainties in Table~\ref{tab:Error}. Consequently, as \pip~ and \citet{cowperley} can be considered independent analyses, we conclude that the algorithm does not contain any visible bias.

\begin{deluxetable}{ccc}[htbp]
    \label{res_table}
    \tablewidth{0.99\textwidth}
    \tablecaption{Temporal properties of the Cow.$^a$} 
    \tablehead{ \colhead{Quantity} & \colhead{Value}  & \colhead{Definition}}
    \startdata
        $\Delta m_g$ & $3.5 \pm 0.5$\,mag & Decline after 15 days rate for the g band \\
        $\Delta m_B$ & $3.5 \pm 0.5$\,mag & Decline after 15 days rate for the $B$ band \\
        $t$          & $30$~days & Time for the decay rate to stabilize for both bands \\
        $\tau_f$     & 0.10\,mag\,day$^{-1}$ & Final decay rate \\
        $\tau_{1w}$  & 0.2--0.4\,mag\,day$^{-1}$ & Decay rates during the first week \\
        $t_{1/2}$    & $3 \pm 0.5$\,days & Time to reach half the peak luminosity in both bands \\
    \enddata
    $^a${Derived with the data produced by {\tt Muphoten}. All of the values are consistent with the published literature on this transient.}
\end{deluxetable}

We also derived some basic timing properties of the Cow with the results produced by \pip~. For both light curves, we interpolated the points obtained from the images we analyzed with splines. The results are shown in cyan in Figure~\ref{lccow}. We computed the decline in magnitude between the peak and 15\,days after in the $g$ band and found $\Delta m \approx 3.5 \pm 0.5$\,mag, the same as for the $B$ band. 
We also evaluated the transient's decay rate based on these results and found a rapid decay during the first week, followed by stabilization to a 0.10\,mag\,day$^{-1}$ rate after 30\,days. These two results are consistent with those found by \citet{cowprentice}. Based on the interpolations, we found a delay of $\sim 3 \pm 0.5$\,days in both bands for the luminosity to decay to half its peak value. This agrees with the value presented by \citet{cowperley}. We summarize all these properties in Table~\ref{res_table}.
Hence, not only are the data produced by our code consistent with independent measurements by other teams and with other telescopes, but we are also able to derive some physical properties of the transient for its characterization.

\section{\textbf{Conclusion}}
\label{ccl}
In this paper, we presented a general method to analyze images of rapidly evolving transients produced by telescopes with different characteristics (depth, bands, location). It estimates the background level, performs the calibration between multiple telescopes using different sets of bandpasses, and measures the transient magnitude in a subtracted image. We implemented two vetoes to check the quality of the calibration and whether the image has issues. The pipeline was tested on the transient SN\,2018cow, which was monitored by various telescopes. The results obtained for all telescopes and all bands are in very good agreement with the published light curves. We also retrieved some timing properties of the transient, giving us confidence in our results.
{\tt Muphoten} can process upcoming images from the GRANDMA network telescopes to identify interesting objects quickly for photometric follow-up observations.

Future developments in the pipeline include the implementation of new methods for estimating the limiting magnitudes. In particular, a recent observing campaign performed by GRANDMA demonstrated the necessity of having more precise limiting-magnitude measurements for offline analysis.


This work is publicly available at \href{https://gitlab.in2p3.fr/icare/MUPHOTEN}{https://gitlab.in2p3.fr/icare/MUPHOTEN}.

M.~W. Coughlin acknowledges support from the National Science Foundation with grant numbers PHY-2010970 and OAC-2117997.
IRiS has received funding from the Excellence Initiative of Aix-Marseille University -- A*MIDEX, a French ``Investissement d'Avenir" programme (ANR-11-LABX-0060 -- OCEVU and AMX-19-IET-008 -- IPhU.
This research made use of Photutils, an Astropy package for detection and photometry of astronomical sources (Bradley et al. 2020). The Liverpool Telescope is operated on the island of La Palma by Liverpool John Moores University in the Spanish Observatorio del Roque de los Muchachos of the Instituto de Astrofisica de Canarias with financial support from the UK Science and Technology Facilities Council. This project has received financial support from the CNRS through the MITI interdisciplinary programs. A.V.F.'s group is grateful for support from the Christopher R. Redlich Fund, the TABASGO Foundation, and the U.C. Berkeley Miller Institute for Basic Research in Science (where A.V.F. was a Miller Senior Fellow).  KAIT and its ongoing operation were made possible by donations from Sun Microsystems, Inc., the Hewlett-Packard Company, AutoScope Corporation, Lick Observatory, the U.S. National Science Foundation, the University of California, the Sylvia and Jim Katzman Foundation, and the TABASGO Foundation. Research at Lick Observatory is partially supported by a generous gift from Google. TAROT was built with the support of the Institut National des Sciences de l’Univers, CNRS, France. TAROT is funded by the CNES, and we acknowledge the help of the technical staff of the Observatoire de Haute Provence, OSU Pytheas.

\software{Sextractor \citep{sex}, Swarp \citep{swarp}, PSFex \citep{psfex}, Scamp \citep{scamp}, photutils \citep{phot}, astropy \citep{2022ApJ...935..167A}, HOTPANTS \citep{hotpants}, astroquery \citep{astroquery} }

\section{Conflict of Interest}
The authors have no relevant financial interests to disclose.
The authors P. A. Duverne, S. Antier, S. Basa, M. W. Coughlin, A. Klotz, and P. Hello declare that they are members of the GRANDMA collaboration.



\section{Availability of Data and Materials}
The data from the GRANDMA collaboration can be shared on reasonable request.
Contact Pierre-Alexandre Duverne to have access to the data used for this work.

\bibliography{references}

\end{document}